\documentclass{CEAS_EuroGNC_2026}

\usepackage[utf8]{inputenc}

\usepackage{graphicx}
\usepackage{amsmath}
\usepackage[version=4]{mhchem}
\usepackage{siunitx}
\usepackage{array}
\usepackage{longtable}
\usepackage{framed}
\usepackage[most]{tcolorbox}
\newtcolorbox{tcbdoublebox}[1][]{%
  enhanced jigsaw,
  sharp corners,
  colback=white,
  borderline={1pt}{-2pt}{black},
  fontupper={\setlength{\parindent}{20pt}},
  #1
}
\usepackage{comment}
\usepackage{subcaption}         
\usepackage{color}		        
\usepackage{booktabs} 
\usepackage{multirow} 
\usepackage{float}  
\usepackage{xcolor}  
\graphicspath{{Figures/}}

\title{Safe Deep Reinforcement Learning for Spacecraft Reorientation with Pointing Keep-Out Constraint}
\papernumber{038}

\ifpdf
\hypersetup{pdfinfo={
Title={Safe Deep Reinforcement Learning for Spacecraft Reorientation with Pointing Keep-Out Constraint},
Author={Juntang Yang},
Keywords={keyword 1, keyword 2, keyword 3, keyword 4} 
}}
\fi

\begin{document}
\thispagestyle{fancy}

\maketitle
\pagestyle{empty} 

\begin{authorList}{4cm} 
\addAuthor{Juntang Yang \orcidlink{0000-0002-7762-3113}}{\textit{\textcolor{gray}{}}Postdoc, University of Würzburg \rorlink{https://ror.org/00fbnyb24}, Würzburg, Germany. \emailAddress{juntang.yang@uni-wuerzburg.de}}
\addAuthor{Mohamed Khalil Ben-Larbi \orcidlink{0000-0002-8390-9302}}{\textit{\textcolor{gray}{}}Professor, University of Würzburg~\rorlink{https://ror.org/00fbnyb24}, Würzburg, Germany. \emailAddress{khalil.ben-larbi@uni-wuerzburg.de}}

\end{authorList}
\justifying

\begin{abstract}
This paper implements deep reinforcement learning (DRL) with a safety filter for spacecraft reorientation control with a single pointing keep-out zone. A new state space representation is designed which includes a compact representation of the attitude constraint zone. A reward function is formulated to achieve the control objective while enforcing the attitude constraint. The soft actor-critic (SAC) algorithm is adopted to handle continuous state and action space. A curriculum learning approach is implemented for agent training. To guarantee the compliance of the attitude constraint, a control barrier function (CBF)-based safety filter is implemented for agent deployment. Simulation results demonstrate the effectiveness of the proposed state space presentation and the designed reward function. Monte Carlo simulations underscore that reward shaping alone cannot guarantee the safety during reorientation maneuver. In contrast, with the CBF-based safety filter, the constraint can be guaranteed during maneuvers.
\end{abstract}

\keywords{Deep Reinforcement Learning, Safe Reinforcement Learning, Spacecraft Attitude Control, Pointing Constraint, Safety Filter, Control Barrier Function}

\color{black}
\section*{Nomenclature}
{\renewcommand\arraystretch{1.0}
\noindent\begin{longtable*}{@{}l @{\quad=\quad} l@{}}
$A_3$ & 3-by-3 submatrix  of $M_F$\\
$h$ & Control barrier function for $\kappa$\\
$\mathcal{H}$ & Inner constraint set\\
$\mathcal{H}^{\Delta}$ & Subset of $\mathcal{H}$ with margin $\Delta$\\
$\mathbb{H}$ & Set of quaternions\\
$I$ &  Moment of inertia of spacecraft\\
$I_3$ & 3-by-3 identity matrix\\
$M_2^+, M_2^-,M_3^+, M_3^-$ & Constants related to $\ddot{\kappa}$\\ 
$M_F$ & Matrix for formulation of keep-out zone constraint\\
$\bar{n}_F$ & Unit central direction vector of keep-out zone\\
$P_\text{f-zone}$ & Keep-out zone penalty\\
$p_h$ & Polynomial as upper bound evolution of $h$\\
$p_\kappa$ & Polynomial as upper bound evolution of $\kappa$\\
$\bm{q}$ & Quaternion\\
$\bm{q}^*$ & Conjugate of quaternion\\
$\bm{q}_d$ & Desired attitude in unit quaternion\\
$\bm{q}_e$ & Relative attitude in unit quaternion\\
$q_0$ & Scalar part of quaternion $\bm{q}$\\
$q_{e0}$ & Scalar part of quaternion $\bm{q}_e$\\
$\bar{q}$ & Vector part of quaternion $\bm{q}$\\
$\mathcal{Q}$ & Safe set\\
$\mathcal{Q}^{\delta}$ & Subset of $\mathcal{Q}$ with margin $\delta$\\
$r, r_1$ & Reward functions\\
$\bar{r}_F$ & Unit boresight vector of instrument\\
$\mathbb{R}$ & Set of real numbers\\
$s$ & State observation\\
$\mathbb{S}^3$ & Set of unit quaternions\\
$t$ & time\\
$t_k$ & Current time step\\
$t_{k-1}$ & Previous time step\\
$T$ & Time step of discretization\\
$\mathbb{U}$ & Set of allowable control inputs\\
$\mathbb{U}_z$ & Set of guaranteed safe control inputs\\
$\bar{x}$ & Full state, $\bar{x} = (\bm{q}, \bar{\omega})$\\
$\mathcal{Z}$ & Robust inner constraint set\\
$\alpha, \beta$ & Positive constants for keep-out zone penalty\\
$\delta$ & Margin (small and positive) for $\mathcal{Q}^{\delta}$\\
$\Delta$ & Margin (small and positive) for $\mathcal{H}^{\Delta}$\\
$\theta$ & Angle between boresight vector $\bar{r}_F$ and avoiding direction $\bar{n}_F$\\
$\theta_F$ & Half angle of keep-out zone\\
$\theta_\text{margin}$ & Safety margin angle, $\theta_\text{margin} = \theta - \theta_F$\\
$\kappa$ & Keep-out zone constraint function\\
$\mu$ & Parameter for control barrier function $h$\\
$\bar{\tau}$ & Control torque\\
$\bar{\tau}_\text{max}$ & Max available control torque\\
$\bar{\tau}_\text{RL}$ & Control torque output by RL agent\\
$\bar{\tau}_\text{safe}$ & Control torque output by safety filter\\
$\bar{\omega}$ & Angular rate\\
$\bar{\omega}^{\times}$ & Skew matrix defined based on $\bar{\omega}$\\
$\bm{\omega}$ & Quaternion form of $\bar{\omega}$\\
$\phi$ & Attitude error angle, $\phi = 2\arccos (q_{e0})$\\
$\psi$ & Certain component of $\ddot{\kappa}$ (2nd time derivative of $\kappa$) under no disturbances\\
$\Delta\bar{n}_F$ & Unit relative avoiding direction, $    \Delta \bar{n}_F = \frac{\bar{n}_F - \bar{r}_F}{\|\bar{n}_F - \bar{r}_F\|}$\\
$\Delta t$ & Arbitrary positive time increment\\
$\Delta\bar{\tau}$ & Change of control torque, $\Delta \bar{\tau}(t_k) = \bar{\tau}(t_k) - \bar{\tau}(t_{k-1})$\\
Superscript $B$ & For variable expressed in body frame\\
Superscript $I$ & For variable expressed in inertial frame\\
$\otimes$ & Quaternion multiplication\\
\end{longtable*}}

\color{black}
\section{Introduction}
Spacecraft reorientation maneuvers are often required to comply with pointing constraints to prevent sensitive onboard instruments (e.g., optical cameras) from exposure to bright celestial objects like the Sun. It is challenging to control spacecraft reorientation under such constraints. \textcolor{black}{There are different approaches for spacecraft constrained reorientation, examples of which include attitude planning methods~\cite{feron2001randomized, kjellberg2013discretized}, nonlinear model predictive control (MPC)-based methods~\cite{gupta2015constrained}, and artificial potential field (APF)-based controllers~\cite{lee2014feedback, yang2021potential}. Attitude planning and nonlinear MPC-based methods are computationally intensive and therefore challenging for real-time application. Despite their computational efficiency, APF-based controllers usually suffer from becoming trapped in local minima.}

Deep reinforcement learning (DRL) offers a promising alternative by combining the computational efficiency of a trained policy with the ability to handle complex, nonlinear dynamics. This makes it well suitable for onboard constrained attitude control. Although initial training is resource-intensive, the resulting agent can generate optimal control commands in real time based on state observations. While DRL has been successfully applied to unconstrained attitude control~\cite{elkins2022bridging, gao2020satellite, Djebko2023learning, oghim2025deep}, its extension to pointing-constrained scenarios remains limited. Recent studies have begun to address this gap. For instance, Jiang et al.~\cite{jiang2023spacecraft} used a Deep Q-Network (DQN) for attitude maneuver planning under forbidden and boundary constraints with a discretized action space, which potentially limits control precision in continuous dynamics. Cai et al.~\cite{cai2024reinforcement} applied the Deep Deterministic Policy Gradient (DDPG) algorithm to a formation flying problem under multiple constraints; however, their state-space design lacked explicit information about the constraint zones, limiting its adaptability. 

A primary challenge in deploying RL for safety-critical systems is the lack of safety guarantees. To address this, the field of safe RL has developed various approaches to ensure operational safety~\cite{gu2024review}. One prominent strategy involves augmenting a performance-driven RL policy with a separate safety filter (SF)~\cite{wabersich2021predictive}. At runtime, the safety filter checks the actions proposed by the agent. If an action is deemed safe, it is executed; if not, the filter overrides it with a safe alternative. While safe RL has seen application in domains like autonomous driving~\cite{isele2018safe}, robotics~\cite{garcia2020teaching}, and spacecraft tasking~\cite{nazmy2022shielded,reed2024shielded}, its potential for ensuring safety in constrained spacecraft attitude control remains largely unexplored.

Motivated by these research gaps, this paper implements DRL for spacecraft reorientation control with a single pointing keep-out constraint in the framework of safe RL. The soft actor-critic (SAC) algorithm~\cite{haarnoja2018soft}  is adopted to handle continuous state and action spaces. The core of our approach includes a state space designed with an explicit and compact representation of the attitude constraint zone, and a tailored reward function to encourage constraint-compliant behavior. To improve training efficiency, a curriculum learning strategy~\cite{narvekar2020curriculum,gupta2022extending} is employed, which progressively increases task difficulty. Crucially, to ensure operational safety, a control barrier function (CBF)-based safety filter~\cite{breeden2023autonomous} is incorporated during agent deployment to guarantee that all maneuvers avoid the keep-out zone. 

The remainder of this paper is organized as follows: Section 2 details the methodology, Section 3 presents and discusses the simulation results, and Section 4 concludes this paper.

\section{Preliminaries}
\subsection{Rotational kinematics and dynamics}
In this paper the set of quaternions is denoted by $\mathbb{H}$ and the set of unit quaternions is denoted by $\mathbb{S}^3$.\\
The rotational kinematic and dynamic equations of a rigid spacecraft are described as~\cite{markley2014fundamentals}
\begin{equation}
\label{eq:kinematics_q}
    \dot{\bm{q}} = \frac{1}{2}\bm{q}\otimes\bm{\omega}
\end{equation}
\begin{equation}
\label{eq:dynamics}
    I\dot{\bar{\omega}} =  - \bar{\omega}^{\times} I \bar{\omega} + \bar{\tau}
\end{equation}
where $\bm{q} = \left[q_0, \bar{q}^T\right]^T \in \mathbb{S}^3$ is the attitude of spacecraft in unit quaternion with $q_0 \in \mathbb{R}$ and $\bar{q} = [q_1, q_2, q_3]^T \in \mathbb{R}^{3}$ as the scalar and vector parts, respectively, $\bm{\omega} = \left[0, \bar{\omega}^T\right]^T \in \mathbb{H}$ is the vector quaternion form of the angular velocity $\bar{\omega} = [\omega_1, \omega_2, \omega_3]^T\in \mathbb{R}^{3}$,  \(I \in \mathbb{R}^{3 \times 3}\) is the moment of inertia of the spacecraft, \(\bar{\tau} = [\tau_1, \tau_2, \tau_3]^T \in \mathbb{R}^{3}\)  is the control torque. Note that both the angular velocity $\bar{\omega}$ and the control torque $\bar{\tau}$ are expressed in the body frame. The skew matrix \(\bar{\omega}^{\times} \in \mathbb{R}^{3 \times 3}\) is defined as
\begin{equation}
\bar{\omega}^{\times}  = 
\left[\begin{matrix}
0 & -\omega_3 & \omega_2 \\
\omega_3 & 0 & -\omega_1 \\
-\omega_2 & \omega_1 & 0
\end{matrix}\right]
\end{equation}
$\otimes$ denotes quaternion multiplication. Given two quaternions $\bm{a} = \left[a_0,~\bar{a}^T\right]^T \in \mathbb{H}$ and $\bm{b} = \left[b_0,~\bar{b}^T\right]^T  \in \mathbb{H}$, the quaternion multiplication between $\bm{a}$ and $\bm{b}$ is defined as
\[
\bm{a}\otimes\bm{b} = \left[\begin{array}{c}
a_0 b_0-\bar{a}\cdot\bar{b}\\
a_0 \bar{b} + b_0 \bar{a} + \bar{a} \times \bar{b}\\
\end{array}\right] 
\]
The conjugate of quaternion $\bm{a}$ is defined as
\[
\bm{a}^* = \left[a_0,~-\bar{a}^T\right]^T  
\]

\subsection{Pointing keep-out zone}
\begin{figure}[h]
\centering
\includegraphics[width=0.618\textwidth]
{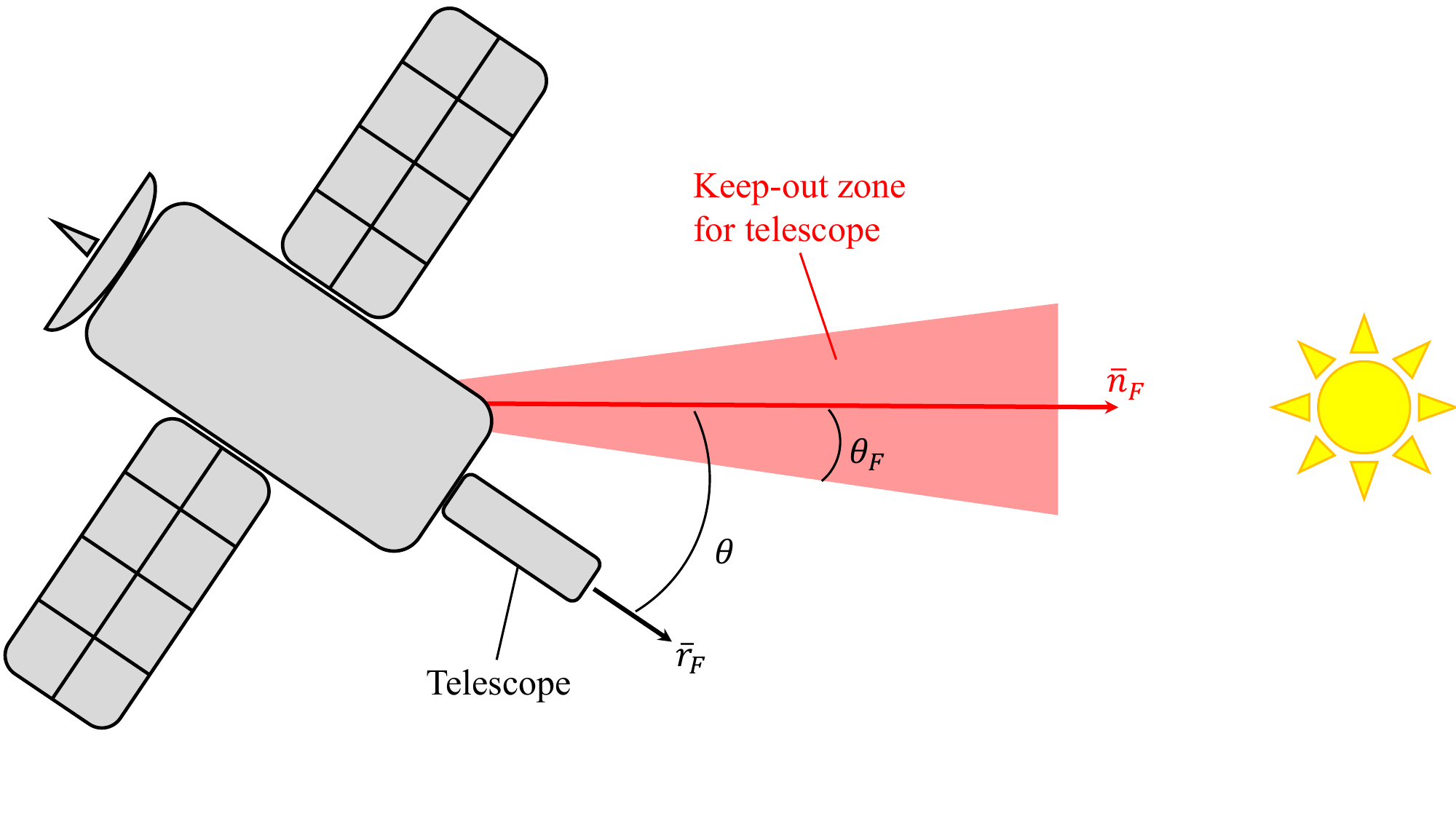}
\caption{Spacecraft sketch with keep-out zone for telescope.}
\label{fig:attitude_constraint_zones}
\end{figure}

A pointing keep-out zone (or forbidden zone) defines an inertial direction that must be avoided by sensitive onboard payloads like telescopes. As shown in Fig.~\ref{fig:attitude_constraint_zones}, the constraint is geometrically represented by a cone, parameterized by its central direction vector $\bar{n}_F$ and half-angle $\theta_F$, which the telescope's boresight vector $\bar{r}_F$ is forbidden to enter. Note that here both $\bar{n}_F$ and $\bar{r}_F$ are unit vectors. 

The mathematical formulation of the requirement of the forbidden zone is written as
\begin{equation}
	\label{eq:requirement_f}
	\bar{r}_F \cdot \bar{n}_F - \cos \theta_F < 0
\end{equation}
which requires the angle between $\bar{r}_F$ and $\bar{n}_F$ (illustrated as $\theta$ in Fig.~\ref{fig:attitude_constraint_zones}) to be larger than $\theta_F$. 

The formulation in Eq.~(\ref{eq:requirement_f}) can be rewritten as~\cite{Lee.2014feedback}
\begin{equation}
	\label{eq:f_zone}
	\bm{q}^{T} M_F \bm{q} < 0		
\end{equation}
with
\begin{equation}
\label{eq:M_F}
M_F = \begin{bmatrix}
		\bar{r}_F^B \cdot \bar{n}_F^I - \cos \theta_F & \left(\bar{r}_F^B \times \bar{n}_F^I\right)^T \\
		\bar{r}_F^B \times \bar{n}_F^I & A_3
		\end{bmatrix}
\end{equation}
where $A_3 = \bar{r}_F^B \left(\bar{n}_F^I\right)^T + \bar{n}_F^I \left(\bar{r}_F^B\right)^T - \left(\bar{r}_F^B \cdot \bar{n}_F^I + \cos \theta_F\right)I_{3}$ with $I_{3} \in \mathbb{R}^{3 \times 3}$ being an identity matrix. The superscripts $B$ and $I$ indicate variables expressed in the body frame and the inertial frame, respectively.

\subsection{Reinforcement learning}

Reinforcement learning (RL) is a machine learning paradigm concerned with how an agent ought to take actions in an environment so as to maximize a notion of cumulative reward~\cite{sutton2018reinforcement, arulkumaran2017deep}. In this learning process, the agent interacts with its environment iteratively. Under the assumption of the Markov property, RL problems are conventionally framed as Markov decision process (MDP). An MDP is formally defined by the tuple~$\left(S,A,R,P,\gamma\right)$, where $S$ and $A$ are the state and action spaces, respectively. The function $R: S \times A \rightarrow \mathbb{R}$ specifies the reward function, $P$ defines the state transition function, and $\gamma \in [0,1)$ is the discount factor. The agent's learning cycle consists of selecting an action $a \in A$ based on the observation of the current state $s \in S$ which results in a transition to a successor state $s^{'}$ and the receipt of a reward $R(s,a)$. The objective in RL is to identify a policy that maximizes the expected cumulative discounted return over the decision horizon.\\
The RL framework for spacecraft reorientation control subject to a single pointing keep-out constraint is formalized as follows.

State space: The state observation, $s(t_k)$, is defined as 
\begin{equation}
\label{eq:state_space}
    s(t_k) = \left[\bm{q}_{e}(t_k), \bar{\omega}(t_k), \bar{r}^B_{F}(t_k), \theta_{\text{margin}}(t_k), \theta(t_k), \Delta \bar{n}^B_{F}(t_k), q_{e0}(t_{k-1})\right]
\end{equation}
where $t_k$ and $t_{k-1}$ indicate the current and previous time step, respectively. The state observation $s(t_k)$ contains the following information: the attitude error $\bm{q}_e(t_k)$ (calculated as $\bm{q}_e(t_k) =\bm{q}_d^* \otimes \bm{q}(t_k)$ with $\bm{q}_d^*$ being the conjugate of the desired attitude $\bm{q}_d$), the angular rate $\bar{\omega}(t_k)$, the payload boresight unit vector in the body frame $\bar{r}^B_F(t_k)$, the angle $\theta(t_k)$ between the boresight vector and the avoid vector as illustrated in Fig.~\ref{fig:attitude_constraint_zones}, the safety margin $\theta_{\text{margin}}(t_k) = \theta(t_k) - \theta_F$, the relative direction vector $\Delta \bar{n}^B_F(t_k)$ (expressed in the body frame), and the previous quaternion scalar component $q_{e0}(t_{k-1})$ providing temporal information. Note that $\theta_{\text{margin}}(t_k) > 0$ must be satisfied to ensure the boresight vector stays outside the keep-out zone. 

The relative direction vector $\Delta \bar{n}^B_F(t_k)$ informs the agent of the correct avoidance maneuver direction and is defined as
\begin{equation}
    \Delta \bar{n}^B_F(t_k) = \frac{\bar{n}^B_F(t_k) - \bar{r}^B_F(t_k)}{\|\bar{n}^B_F(t_k) - \bar{r}^B_F(t_k)\|}
\end{equation}
where $\bar{n}^B_F(t_k)$ is the unit vector $\bar{n}_F$ (see Fig.~\ref{fig:attitude_constraint_zones}) expressed in the body frame.

Action space: the action space is defined by the body-frame control torque $\bar{\tau} \in \mathbb{R}^3$, which is normalized to $[-1, 1]$ during training.

Reward function:  The reward function $r(t_k)$ at the current time step is formulated based on the design by Elkins et al.~\cite{elkins2022bridging} as follows:   
\begin{equation}
\label{eq:reward_r}
r(t_k) = 
\begin{cases}
r_{1}(t_k) + 9,	& \phi(t_k) \leq 0.25^{\circ}\\
r_{1}(t_k), 	& \text{otherwise}
\end{cases}
\end{equation}
\begin{equation}
    \label{eq:reward_r1}
r_{1}(t_k) = 
    \begin{cases}
    e^{\frac{-\phi(t_k)}{0.14*2\pi}} - 0.05\frac{\|\bar{\tau}(t_k)\|}{\|\bar{\tau}_{\text{max}}\|} - 0.005\|\Delta \bar{\tau}(t_k)\| - \text{P}_{\text{f-zone}}(t_k),	& q_{e0}(t_k) > q_{e0}(t_{k-1})\\ \\
    e^{\frac{-\phi(t_k)}{0.14*2\pi}} - 0.05\frac{\|\bar{\tau}(t_k)\|}{\|\bar{\tau}_{\text{max}}\|} - 0.005\|\Delta \bar{\tau}(t_k)\| - \text{P}_{\text{f-zone}}(t_k) -1, 	& \text{otherwise}
    \end{cases}
\end{equation}
where $\phi(t_k) = 2\arccos (q_{e0}(t_k))$ is the attitude error angle, $\bar{\tau}_{\text{max}}$ is the max available control torque, and $\Delta \bar{\tau}(t_k) = \bar{\tau}(t_k) - \bar{\tau}(t_{k-1})$ is the change of the torque. The keep-out zone penalty is defined as:
\begin{equation}
\text{P}_{\text{f-zone}}(t_k) = 
\begin{cases}
\beta,	& \theta_{\text{margin}}(t_k) \leq 0\\
\beta e^{-\alpha \theta_{\text{margin}}(t_k)}, 	& \text{otherwise}
\end{cases}
\end{equation}
where $\alpha$ and $\beta$ are positive constants, and $\theta_\text{margin}(t_k)$ is in radians.

\begin{figure}[htbp]
\centering
\begin{subfigure}{0.45\textwidth}
\includegraphics[width=\linewidth]{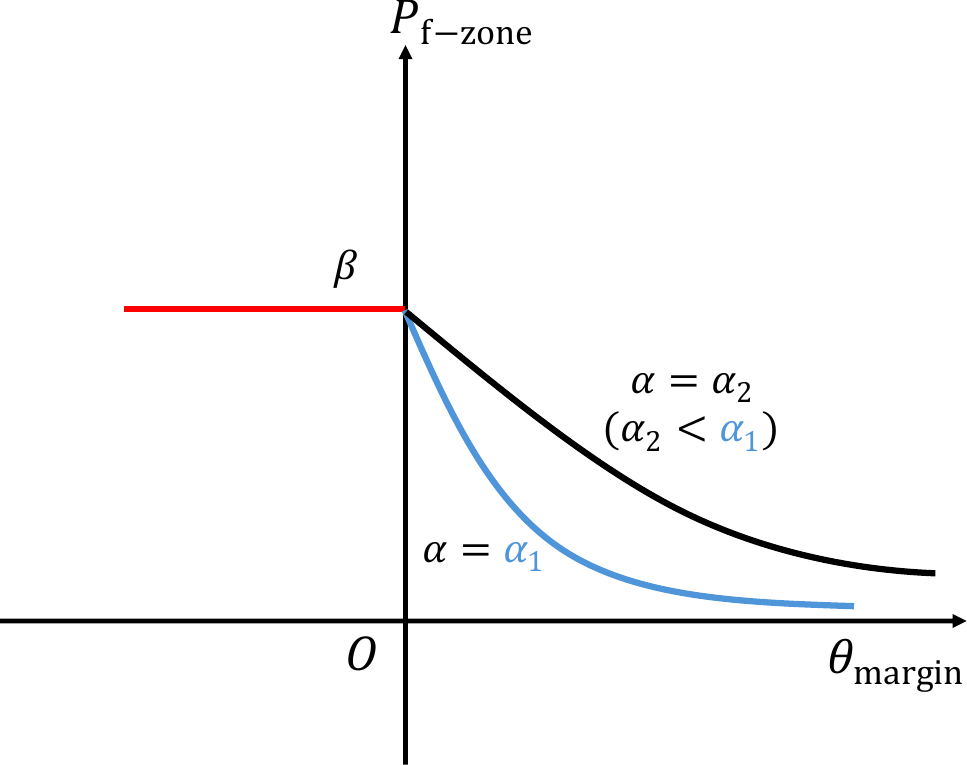}
\caption{}
\label{fig:explanation_P_fzone_a}
\end{subfigure}
\begin{subfigure}{0.35\textwidth}
\includegraphics[width=\linewidth]{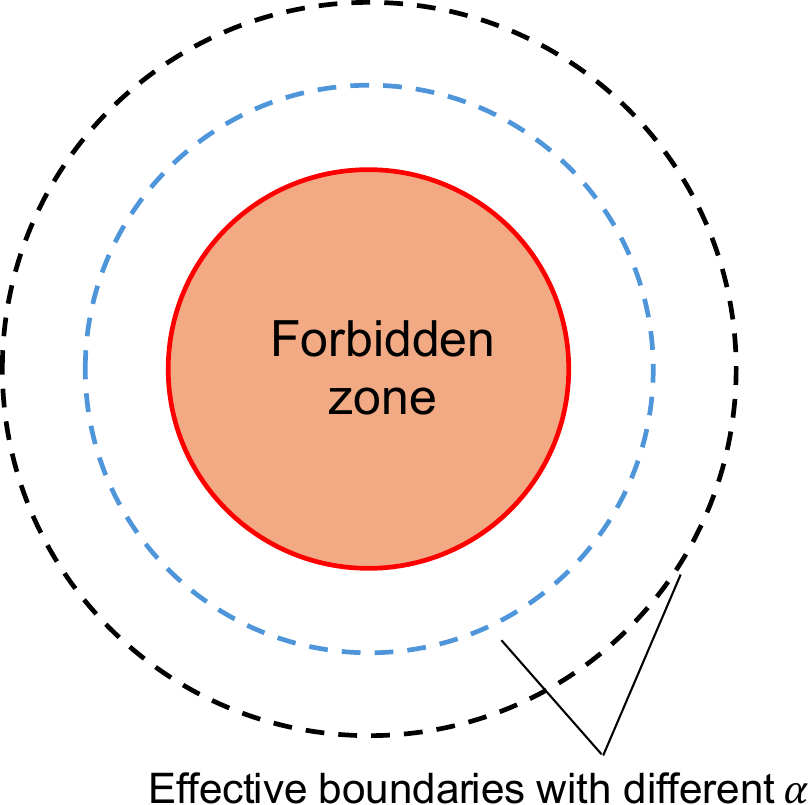}
\caption{}
\label{fig:explanation_P_fzone_b}
\end{subfigure}
\caption{Illustration of keep-out zone penalty $\text{P}_{\text{f-zone}}$} 
\label{fig:explanation_P_fzone}
\end{figure}

\textcolor{black}{The reward function in Eq.~(\ref{eq:reward_r}) is composed of two primary elements: a base reward $r_{1}(t_k)$ and an extra reward of 9 when achieving the desired pointing accuracy of $0.25^\circ$. The base reward $r_{1}(t_k)$ in Eq.~(\ref{eq:reward_r1}) consists of weighted terms promoting attitude convergence, minimizing control effort and its change, and penalizing violations of the keep-out zone constraint via the penalty function $\text{P}_{\text{f-zone}}(t_k)$. Figure~\ref{fig:explanation_P_fzone} visualizes the keep-out zone penalty function with two different $\alpha$ values (see Fig.~\ref{fig:explanation_P_fzone_a}) and illustrates how the parameter $\alpha$ affects the effective boundary (indicated as the dashed lines in Fig.~\ref{fig:explanation_P_fzone_b}), outside which the keep-out zone penalty is below a defined small threshold value and can be ignored.} Increasing $\alpha$ narrows this boundary, concentrating the penalty closer to the zone.

State transition: The state transition is governed by the rotational kinematics and dynamics, as given by Eqs.~(\ref{eq:kinematics_q}) and (\ref{eq:dynamics}).

\subsection{RL with safety filter}
\begin{figure}[htbp]
\begin{subfigure}{0.49\textwidth}
\includegraphics[width=\linewidth]{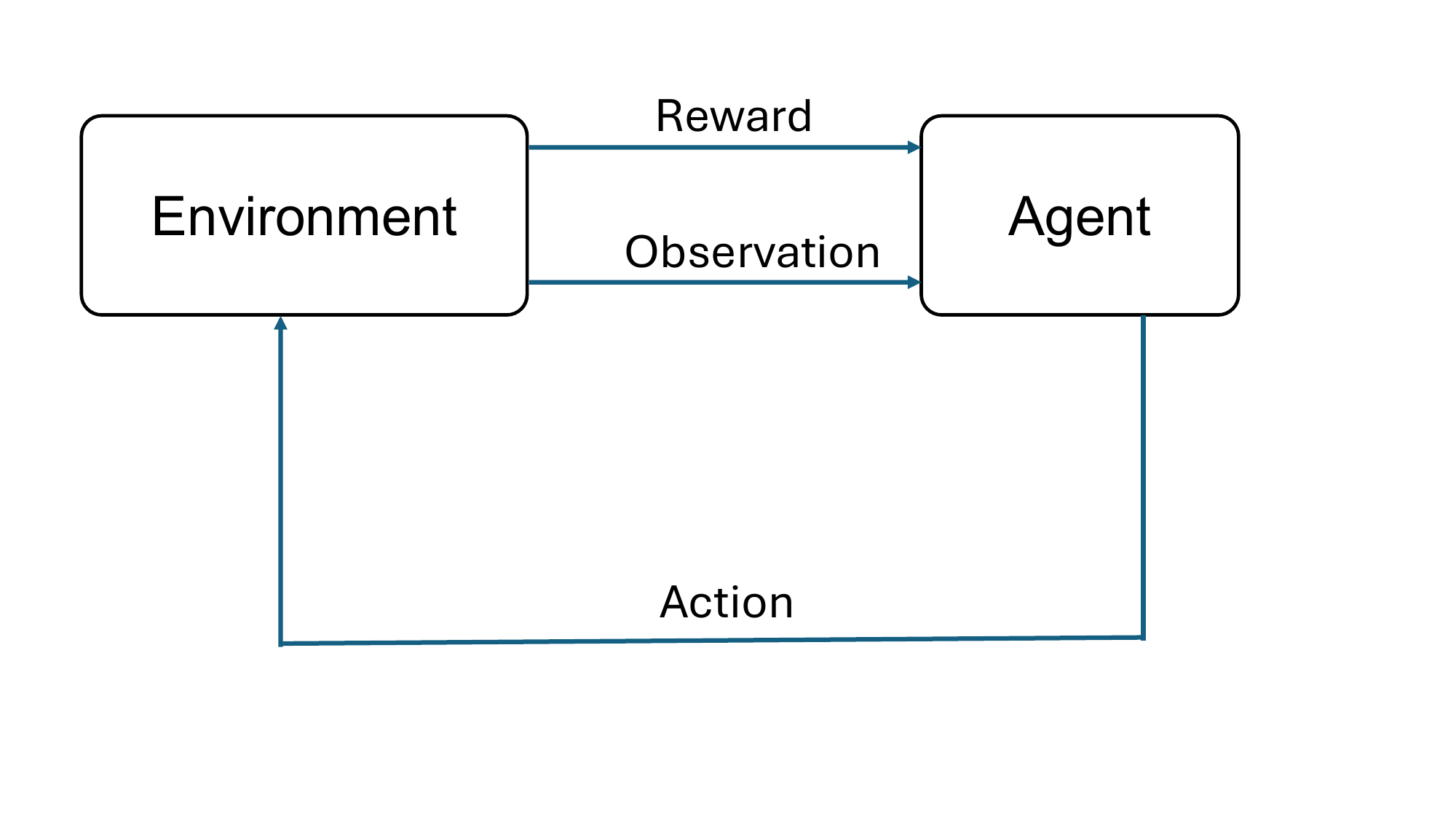}
\caption{Standard RL}
\label{fig:basic_RL}
\end{subfigure}
\begin{subfigure}{0.49\textwidth}
\includegraphics[width=\linewidth]{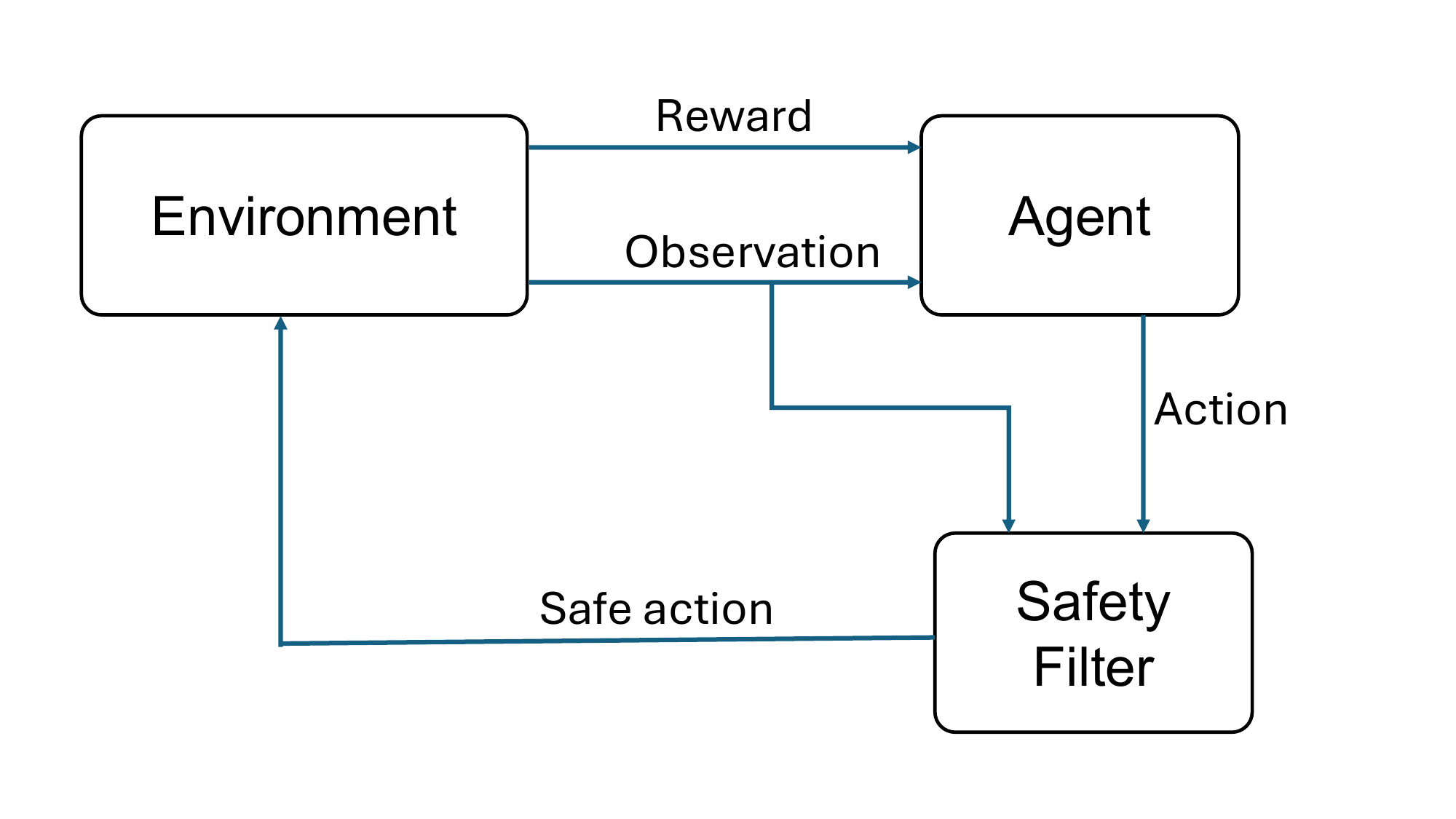}
\caption{RL with safety filter}
\label{fig:Safety_filter_RL}
\end{subfigure}
\caption{Standard RL and RL with safety filter} 
\label{fig:two_RL_frameworks}
\end{figure}

Figure~\ref{fig:two_RL_frameworks} compares a standard RL framework with a safety-augmented variant. The key difference is that the latter interposes a safety filter between the agent and the environment (Fig.~\ref{fig:Safety_filter_RL}) to modify actions for safety, unlike the direct application in the standard framework (Fig.~\ref{fig:basic_RL}).  

In this paper, we implement a control barrier function (CBF)-based safety filter developed by Breeden and Panagou~\cite{breeden2023autonomous}. The basic idea of the safety filter is as follows: at each time step $t_k$, the safety filter finds a control torque $\bar{\tau}_\text{safe}(t_k)$ closest to the agent action, $\bar{\tau}_{RL}(t_k)$, while satisfying the control limit and safety constraints by solving the following optimization problem
\begin{eqnarray} 
\label{eq:opt1}
\bar{\tau}_\text{safe}(t_k) = \arg \min_{\bar{\tau}(t_k) \in \mathbb{U} \cap \mathbb{U}_{z}}& \|\bar{\tau}(t_k) - \bar{\tau}_{RL}(t_k)\|^2
\end{eqnarray}
where $\mathbb{U}$ is the set of allowable control inputs, and $\mathbb{U}_{z}$ is the set of control inputs guaranteeing the compliance of the keep-out zone constraint.

\color{black}
The control set $\mathbb{U}$ is derived from the operational limits of the actuators. The calculation of $\mathbb{U}_{z}$ is briefly explained  in the remainder of this section, with slight adaptation of the notation used in~\cite{breeden2023autonomous}. Note that, for the sake of brevity, the dependency of $\bm{q}$ and $\bar{\omega}$ on the time $t$ is omitted in the following explanation. 

The safe set $\mathcal{Q}(t)$ is defined as 
\begin{equation}
    \mathcal{Q}(t) \triangleq \left\{(\bm{q},\bar{\omega})|\kappa(t,\bm{q})< 0\right\}
\end{equation}
with 
\begin{equation}
\label{eq:kappa}
	\kappa (t, \bm{q}) \triangleq \bm{q}^{T} M_F \bm{q}		
\end{equation}
The states in $\mathcal{Q}$ ensures the satisfaction of the keep-out zone constraint at the current time instant.

Breeden and Panagou~\cite{breeden2023autonomous} utilize the CBF theory to ensure state trajectories always inside the safe set and they choose the CBF for $\kappa$ as
\begin{equation}
\label{eq:CBF}
    h(t, \bm{q}) = \kappa (t, \bm{q}) + \frac{\dot{\kappa}(t, \bm{q})|\dot{\kappa}(t, \bm{q})|}{2\mu}
\end{equation}
with the parameter  $ \mu \in (0, 0.0025]$. Note that the CBF with a larger $\mu$ is less conservative.

The inner constraint set $\mathcal{H} (t)$ for the keep-out zone is defined as
\begin{equation}
    \mathcal{H} (t) \triangleq \left\{(\bm{q},\bar{\omega})|h(t,\bm{q})\leq 0\right\}
\end{equation}
The state trajectories inside $\mathcal{H}$ cannot leave $\mathcal{H}$, i.e., $\mathcal{H}$ is a controlled-invariant set~\cite{breeden2023autonomous}. 

In order to account for input constraints, disturbances, and controller sampling, the robust inner constraint set is defined as~\cite{breeden2023autonomous}
\begin{equation}
    \mathcal{Z}(t) = \mathcal{Q}^{\delta}(t) \cap \mathcal{H}^{\Delta} (t)
\end{equation} 
where $\mathcal{Q}^{\delta}(t)$ and $\mathcal{H}^{\Delta} (t)$ are the subset of $\mathcal{Q}(t)$ with margin ${\delta}$ and the subset of $\mathcal{H}(t)$ with margin ${\Delta}$, respectively, defined as
\begin{equation}
    \label{eq:Q_delta}
    \mathcal{Q}^{\delta}(t) \triangleq \left\{(\bm{q},\bar{\omega})|\kappa(t,\bm{q})\leq -\delta\right\}
\end{equation}
\begin{equation}
    \label{eq:H_Delta}
    \mathcal{H}^{\Delta} (t) \triangleq \left\{(\bm{q},\bar{\omega})|h(t,\bm{q})\leq - \Delta\right\}
\end{equation}
with $\delta$ and $\Delta$ as small positive parameters.

For control based on sampled data, the robust inner constraint set $\mathcal{Z}(t)$ with suitable $\delta$ and $\Delta$ can ensure the safety between time steps~\cite{breeden2023autonomous}. 

The requirements in Eqs.~(\ref{eq:Q_delta}) and~(\ref{eq:H_Delta}) are further transformed as the requirements on the upper bound evolution of $\kappa$  (denoted as $p_\kappa$) and the upper bound evolution of $h$ (denoted as $p_h$), which is expressed as 
\begin{equation}
    p_\kappa(t,\bar{x},\bar{\tau},\Delta t) \leq -\delta
\end{equation}
\begin{equation}
    p_h(t,\bar{x},\bar{\tau},\Delta t) \leq -\Delta  
\end{equation}
with $p_\kappa$ and $p_h$ as polynomials in $\Delta t$ (an arbitrary positive time increment) defined as follows
\begin{equation}
    p_\kappa(t,\bar{x},\bar{\tau},\Delta t) \triangleq \kappa(t,\bar{x}) + \dot{\kappa}(t,\bar{x}) \Delta t + \frac{1}{2}\psi(t,\bar{x},\bar{\tau})\left(\Delta t\right)^2 + \frac{1}{2}M_2^+ \left(\Delta t\right)^2 + \frac{1}{6}M_3^+\left(\Delta t\right)^3
\end{equation}
\begin{equation}
    p_h(t,\bar{x},\bar{\tau},\Delta t) \triangleq p_\kappa(t,\bar{x},\bar{\tau}, \Delta t) + \frac{1}{2\mu} \text{ssq}\left(\dot{\kappa}(t,\bar{x}) \Delta t + \psi(t,\bar{x},\bar{\tau}) \Delta t + M_2^+ \Delta t + \frac{1}{2}M_3^+ \left(\Delta t\right)^2\right)  
\end{equation}
where $\bar{x} = (\bm{q}, \bar{\omega})$, $\psi(t,\bar{x},\bar{\tau})$ is the certain component of $\ddot{\kappa}$ under no disturbances, the constant $M_2^+$ represents the upper bound on the uncertainty in $\ddot{\kappa}$ because of unknown disturbances, the constant $M_3^+$ describes the uncertainty in the evolution of $\psi(t,\bar{x},\bar{\tau})$ between time steps due to both the zero-order-hold (ZOH) sampling and disturbances. $\text{ssq}(\lambda) \triangleq \lambda |\lambda|$ for brevity. Note that the dependency of $\bar{x}$ and $\bar{\tau}$ on the time is omitted for the sake of brevity. 

Based on $p_\kappa$ and $p_h$, $\mathbb{U}_{z}$ in Eq.~(\ref{eq:opt1}) is defined as
\begin{equation}
\label{eq:U_z}
    \mathbb{U}_{z} = \left\{\bar{\tau} \in \mathbb{R}^3 | p_\kappa(t_k,\bar{x},\bar{\tau},T) \leq -\delta~\text{and}~p_h(t_k,\bar{x},\bar{\tau},T) \leq -\Delta\right\}
\end{equation}
where $t_k$ is the sample time and $T$ is the time step of discretization. Note that both $p_\kappa(t_k,\bar{x},\bar{\tau},T) \leq -\delta$ and $p_h(t_k,\bar{x},\bar{\tau},T) \leq -\Delta$ in Eq.~(\ref{eq:U_z}) can be encoded in a quadratic program (QP). As for the calculation of $\mathbb{U}_{z}$, readers are referred to~\cite{breeden2023autonomous} for more details. 

\color{black}
\subsection{Agent training}
This work employs the SAC algorithm~\cite{haarnoja2018soft} for agent training due to its suitability for continuous action spaces and strong exploration capabilities. As an off-policy actor-critic method, SAC offers high sample efficiency. SAC maximizes an entropy-augmented objective, which balances task performance with exploration by encouraging stochastic behavior. The implementation is based on the Stable-Baselines3 library~\cite{raffin2021stable}.

For agent training, the simulation runs with a time step $T=0.1~\text{s}$ over episodes of \SI{100}{s} duration. For attitude regulation scenarios, the initial angular rate is set as zero with small random perturbations. The spacecraft's moment of inertia is fixed at
\[I = \left[\begin{matrix}
60 & 5 & 1 \\
5 & 50 & 2 \\
1 & 2 & 70
\end{matrix}\right] \text{kg}\cdot\text{m}^2\]
The boresight vector is $[1,0,0]^T$. The torque limit for each axis is \SI{2}{Nm}. The keep-out zone penalty parameters are $\beta = 10$, $\alpha = 66$. 

Agent training followed a two-phase curriculum learning strategy. 
In Phase 1, the agent learned a baseline attitude control policy without keep-out zones. The initial attitude error was randomized, with the maximum initial deviation angle progressively increased from $25^{\circ}$ to $180^{\circ}$. The resulting policy and experience replay buffer from this phase were saved for initialization in further training.
In Phase 2, the pre-trained agent was fine-tuned with the keep-out zone penalty activated. For each episode, a single keep-out zone was generated by placing its center vector, $\bar{n}_F$, along the shortest rotation path between the initial attitude (with randomized error between $80^{\circ}$ and $180^{\circ}$ ) and the target. The cone half angle, $\theta_F$, was randomized between $15^{\circ}$ and $30^{\circ}$. This placement ensures the zone intersects the agent's likely path since the policy from Phase 1 generates short-path rotations. The maximum initial deviation angle was again gradually increased from $80^{\circ}$ to $180^{\circ}$.

Agent training was conducted on a desktop PC (Intel i7-14700 KF, 32 GB RAM, NVIDIA RTX 4070 GPU). The default SAC configuration from Stable-Baselines3 was used, which features an MlpPolicy with ReLU activations and a two-layer fully connected network architecture with 256 units per layer. The hyperparameters used for training are provided in Table~\ref{tab:hyperparameters}.

\begin{table}[htbp]
\caption{\label{tab:hyperparameters} SAC hyperparameters for agent training}
\centering
\begin{tabular}{cc}
\hline
\hline
Parameter & Value \\
\hline
Batch size  & 256 \\ 
Buffer size & $10^6$ \\
Discount factor ($\gamma$) & 0.99 \\
Entropy coefficient & Auto \\ 
Learning rate (without F-zone) & 0.0001 \\
Learning rate (with F-zone) & 0.0001\\
Soft update coefficient & 0.005 \\
\hline
\hline
\end{tabular}
\end{table}

\section{Numerical Results}
\begin{figure}[htbp]        
\begin{subfigure}{0.45\textwidth}
\includegraphics[width=\linewidth]{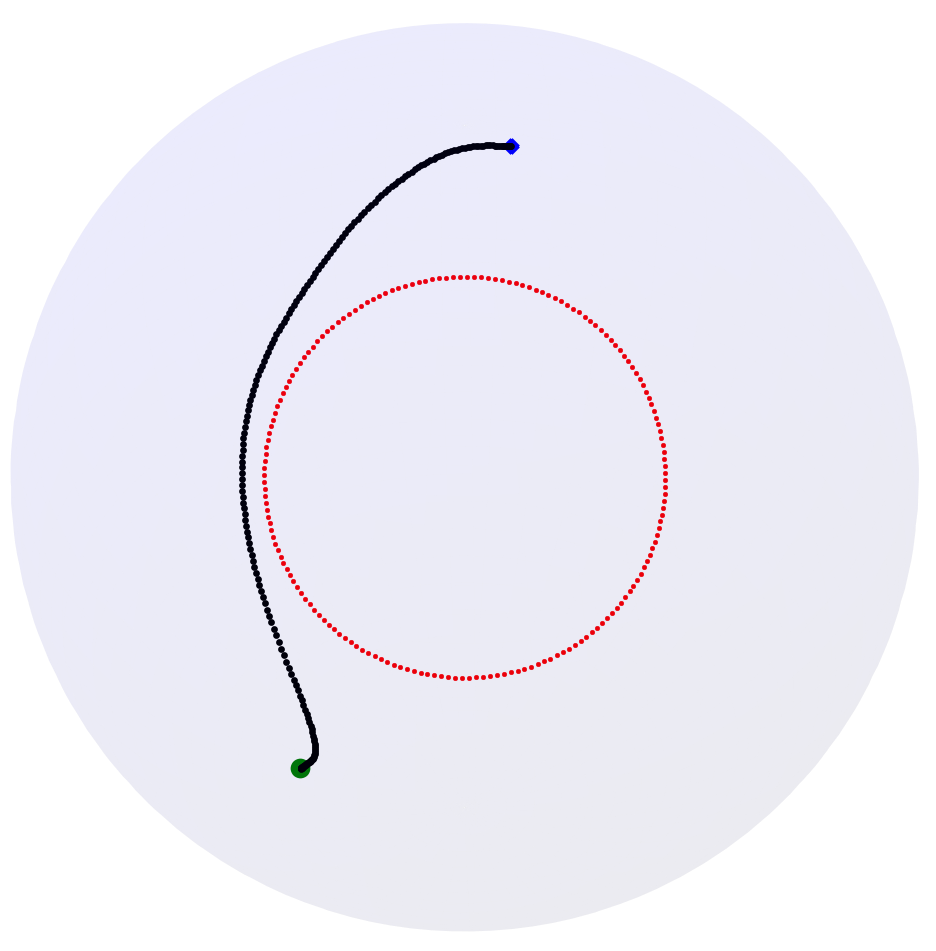}
\caption{Trace of boresight vector on unit sphere}
\label{fig:plot_3d_withFzone}
\end{subfigure}
\begin{subfigure}{0.55\textwidth}
\includegraphics[width=\linewidth]{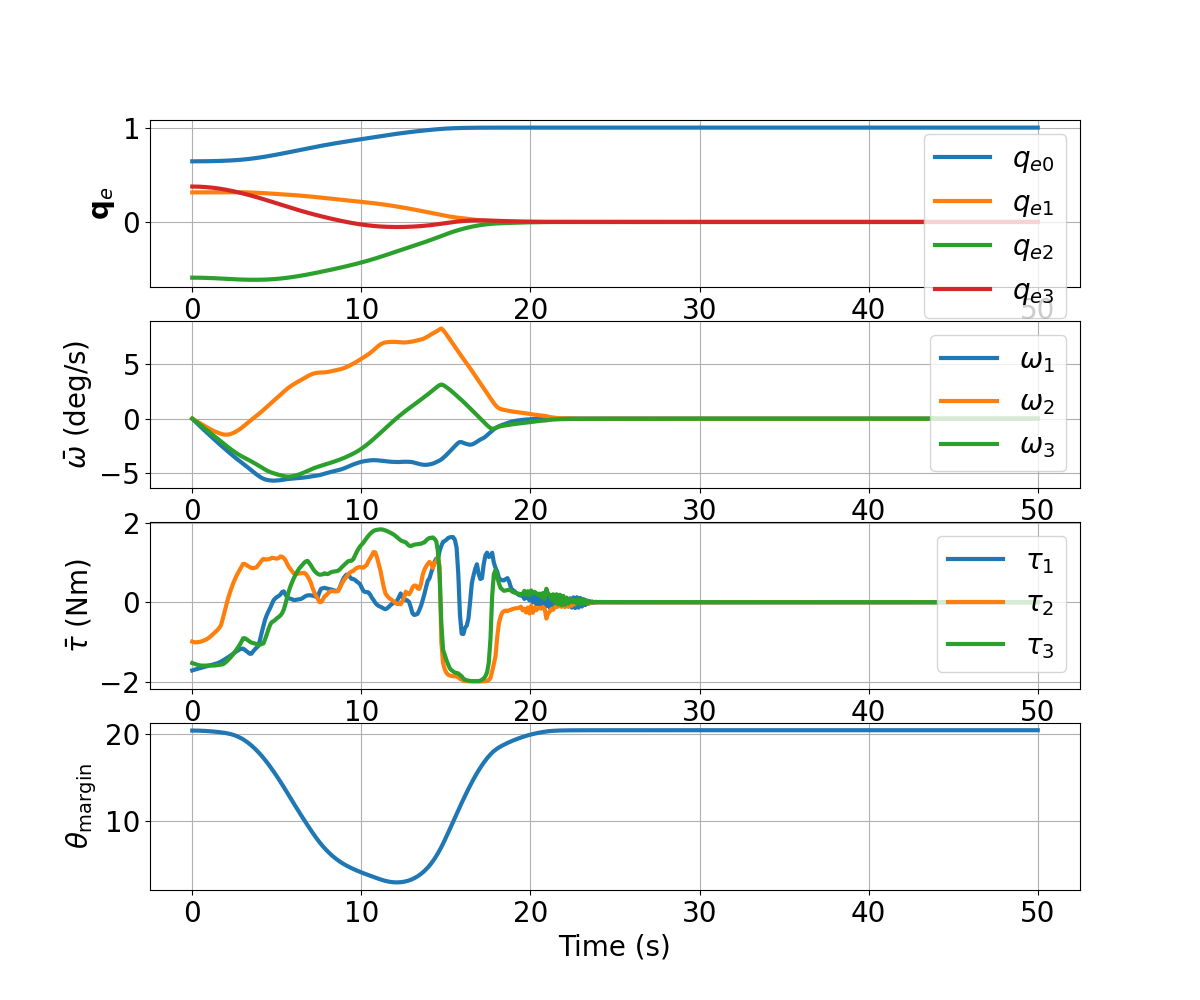}
\caption{Time history of relative attitude, angular velocity, control torque, and $\theta_\text{margin}$}
\label{fig:plot_data_withFzone}
\end{subfigure}
\caption{Example result under agent trained in Phase 2 (with one F-zone)} 
\label{fig:example_result_withFzone}
\end{figure}

A reorientation scenario with an initial deviation angle of $100^{\circ}$ and one F-zone was tested with an agent trained in Phase 2. The parameters are detailed in Table~\ref{tab:data_simulation}. 
\begin{table}[htbp]
\caption{\label{tab:data_simulation} Parameters for example simulation}
\centering
\begin{tabular}{cc}
\hline
\hline
Parameter & Value \\
\hline
Avoid vector ($\bar{n}^I_{F}$) & $[0.703, 0.263, 0.661]^T$ \\
Boresight vector ($\bar{r}^B_{F}$) & $[1,0,0]^T$ \\
Half angle ($\theta_F$) & 25 deg \\ 
Initial relative attitude ($\bm{q}_e$)  & $[0.6428, 0.3138, -0.5892, 0.3757]^T$ \\ 
Initial angular rate ($\bar{\omega}$) & $[-5.7,-1.1,-9.9]^T*10^{-4}~\text{deg/s}$ \\
Target attitude  & $[1,0,0,0]^T$ \\ 
\hline
\hline
\end{tabular}
\end{table}
An example result under the agent is presented in Fig.~\ref{fig:example_result_withFzone}. Figure~\ref{fig:plot_3d_withFzone} shows the boresight vector's trace on the unit sphere. The trace (black) starts at the initial pointing (blue cross) and ends at the target (green point), successfully avoiding the keep-out zone bounded by the red circle. The corresponding time histories in Fig.~\ref{fig:plot_data_withFzone} confirm that the spacecraft achieves the desired attitude as $\bm{q}_e$ converges to the identity quaternion, while maintaining a positive margin angle $\theta_{\text{margin}}$ throughout the maneuver. These results validate that the design of the state space and the reward function is effective for RL-based attitude control with a single pointing keep-out zone.

\begin{figure}[htbp]
\centering
\begin{subfigure}{0.75\textwidth}
\includegraphics[width=\linewidth]{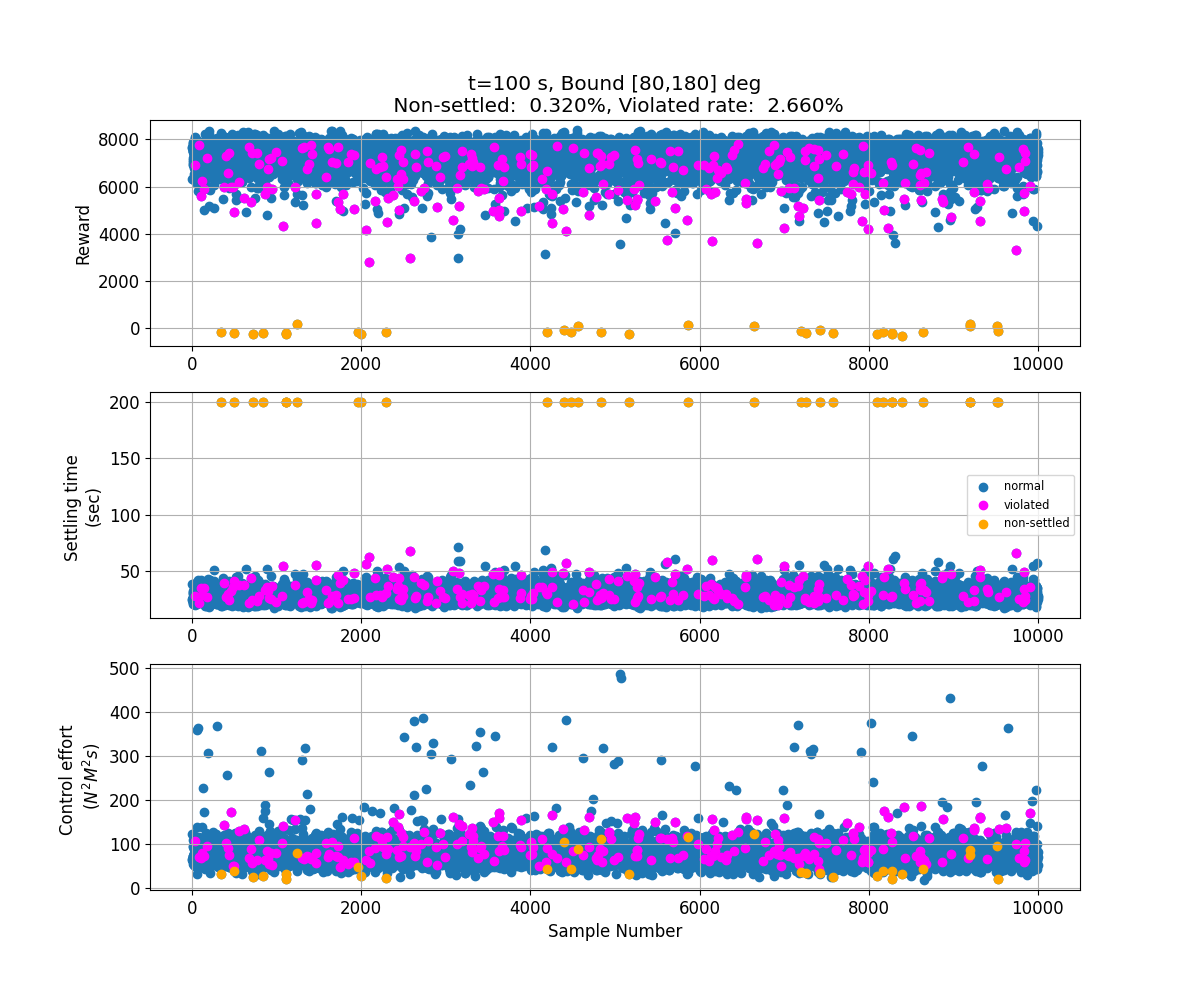}
\caption{Without using safety filter (namely, in standard RL framework)}
\label{fig:plot_monte_carlo_withoutSF}
\end{subfigure}
\begin{subfigure}{0.75\textwidth}
\includegraphics[width=\linewidth]{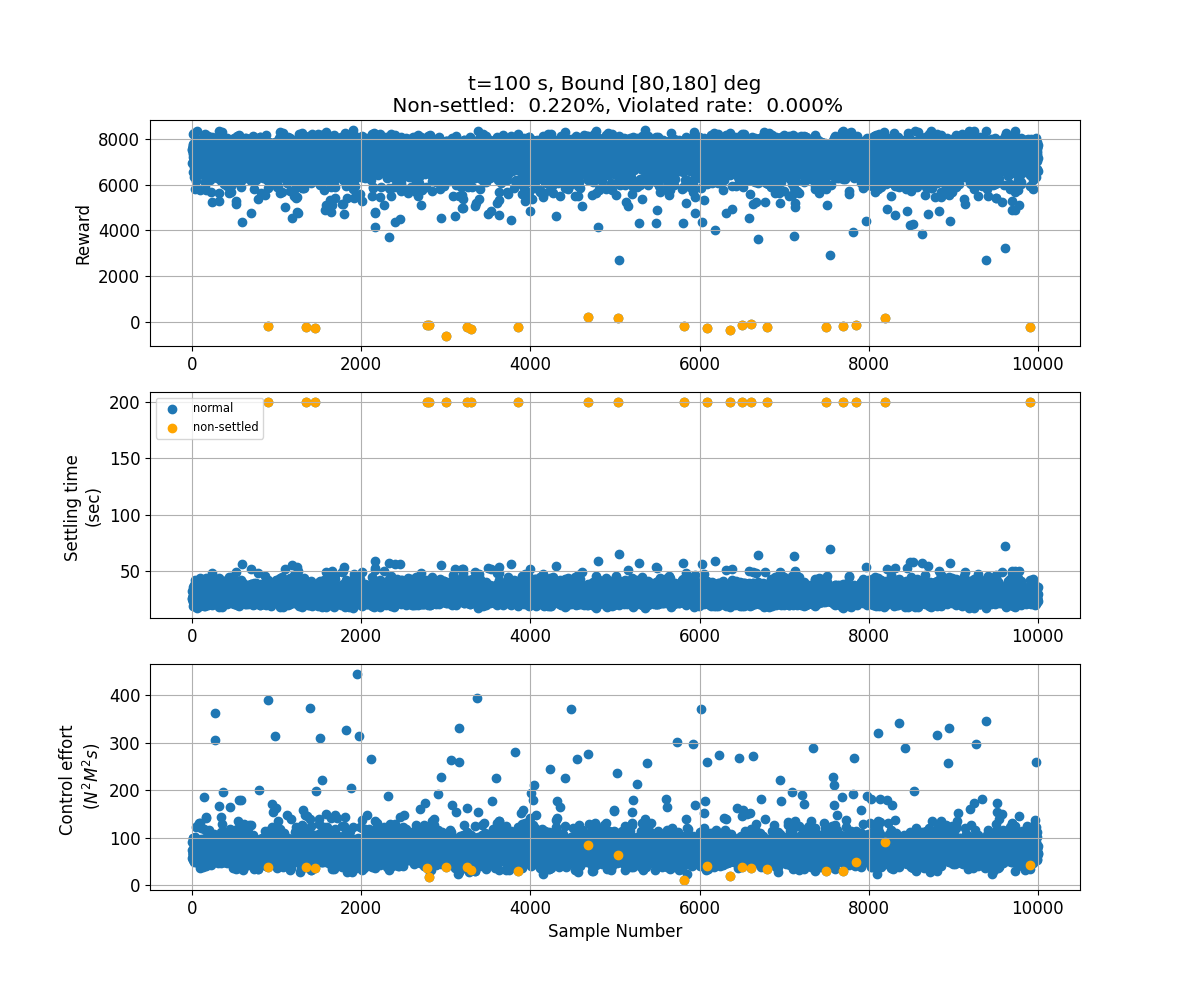}
\caption{Using safety filter (namely, in safe-RL framework)}
\label{fig:plot_monte_carlo_withSF}
\end{subfigure}
\caption{Monte Carlo simulation results (metrics vs sample number) under the best rewarded agent trained in Phase 2 (with one F-zone). The non-settled cases are assigned a settling time of 200 s for visualization.} 
\label{fig:plot_monte_carlo_1_and_2}
\end{figure}

A Monte Carlo analysis comprising 10,000 simulations was conducted to statistically evaluate the performance of the best rewarded agent from Phase 2. Each scenario featured a random initial deviation angle between $80^{\circ}$ and $180^{\circ}$ and an initial angular rate uniformly distributed in the range $[-0.001, 0.001]~^{\circ}$/s. The simulation uses the same time step, duration, and moment of inertia as in the training phase. Each test was evaluated based on the cumulative reward, settling time (\textcolor{black}{defined as the time at which the attitude error enters and subsequently remains within the desired accuracy of 0.25 deg}), control accuracy, and the control effort, defined as \[E\left(t_{end}\right) = \int_{0}^{t_{end}} \|\bar{\tau}\|^2 dt\]
The results of Monte Carlo simulation using the framework of standard RL are summarized in Fig.~\ref{fig:plot_monte_carlo_withoutSF} showing that the agent successfully reached the target orientation while complying with the constraint zone in approximately $97\%$ of cases (blue samples). The failures (about $3\%$) were due to constraint violations (purple samples, $2.66\%$) and failure to converge within the simulation duration (orange samples, $0.32\%$). Note that the latter failures were assigned a settling time of 200 s for visualization. The detailed results of evaluation metrics are listed in Table~\ref{tab:monte_carlo_statistics_comparison} (2nd column).
The insight from the Monte Carlo analysis is that incorporating the keep-out zone penalty into the reward function alone does not guarantee full constraint compliance, as evidenced by the $2.66\%$ violation rate. This underscores the need for additional safety assurances when using the standard RL framework.

\begin{table}[htbp]
\centering
\caption{Results of Monte Carlo simulation}
\begin{tabular}{l|ccc}
\hline
\hline
\multirow{3}{*}{Evaluation metrics} & \multicolumn{3}{c}{Results} \\
\cline{2-4}
                         & Standard RL & RL with SF & RL with SF \\
                         &             & ($\mu = 0.0025$) & ($\mu = 0.0001$)\\                        
\hline
Mean Reward  & $7281.91 \pm 688.85$ & \textcolor{black}{$7232.96 \pm 642.41$} & $6372.58 \pm 1327.31$\\ 
Mean settling time* (sec) & $27.81 \pm 5.24$ & \textcolor{black}{$28.47 \pm 5.30$} & $37.21 \pm 12.36$\\
Mean control effort* ($\text{N}^2\text{m}^2\text{s}$) & $76.02 \pm 25.76$ & \textcolor{black}{$73.31 \pm 24.21$} & $68.65 \pm 18.00$ \\
Mean control accuracy* (deg) & $0.08 \pm 0.04$ & $0.08 \pm 0.04$ & $0.08 \pm 0.04$\\
Rate of non-settled & $0.32\%$ & \textcolor{black}{$0.22\%$} & $0.70\%$\\ 
Rate of violation & $2.66\%$ & $0\%$ & $0\%$\\
\hline
\hline
\end{tabular}
\vspace{0.3cm}
\small
\raggedright
$^*$Non-settled samples are ignored.
\label{tab:monte_carlo_statistics_comparison}
\end{table}

The same agent was used for a Monte Carlo simulation with the CBF-based safety filter in~\cite{breeden2023autonomous} in the loop. The parameters of the safety filter are presented in Table~\ref{tab:parameters_SF}.  
\textcolor{black}{Small $M_2^+$, $M_2^-$, $M_3^+$, and $M_3^-$ are used since no disturbances are considered. For the least conservative CBF of the form in Eq.~(\ref{eq:CBF}), the largest allowable parameter $\mu = 0.0025$ is used~\cite{breeden2023autonomous}. The calculation of $\delta$ and $\Delta$ is based on the setting of $M_2^+$, $M_2^-$, $M_3^+$, $M_3^-$, $\mu$ and the time step $T$ (See Theorem 1 in~\cite{breeden2023autonomous} for more details).} 
\begin{table}[htbp]
\caption{\label{tab:parameters_SF} Parameters of CBF-based safety filter}
\centering
\begin{tabular}{c|c}
\hline
\hline
Parameter & Value \\
\hline
$T$  & $0.1$~s \\
$M_2^+$  & $1.64 \times 10^{-5}$ \\ 
$M_2^-$ & $-1.64 \times 10^{-5}$ \\
$M_3^+$ & $6.2 \times 10^{-4}$ \\
$M_3^-$ & $-6.2 \times 10^{-4}$ \\ 
$\mu$   & $\textcolor{black}{0.0025}$\\
$\delta$ & $\textcolor{black}{3.18} \times 10^{-6}$ \\
$\Delta$ & $\textcolor{black}{3.18} \times 10^{-6}$\\
\hline
\hline
\end{tabular}
\end{table}

\begin{figure}[htbp]   
\begin{subfigure}{0.45\textwidth}
\includegraphics[width=\linewidth]{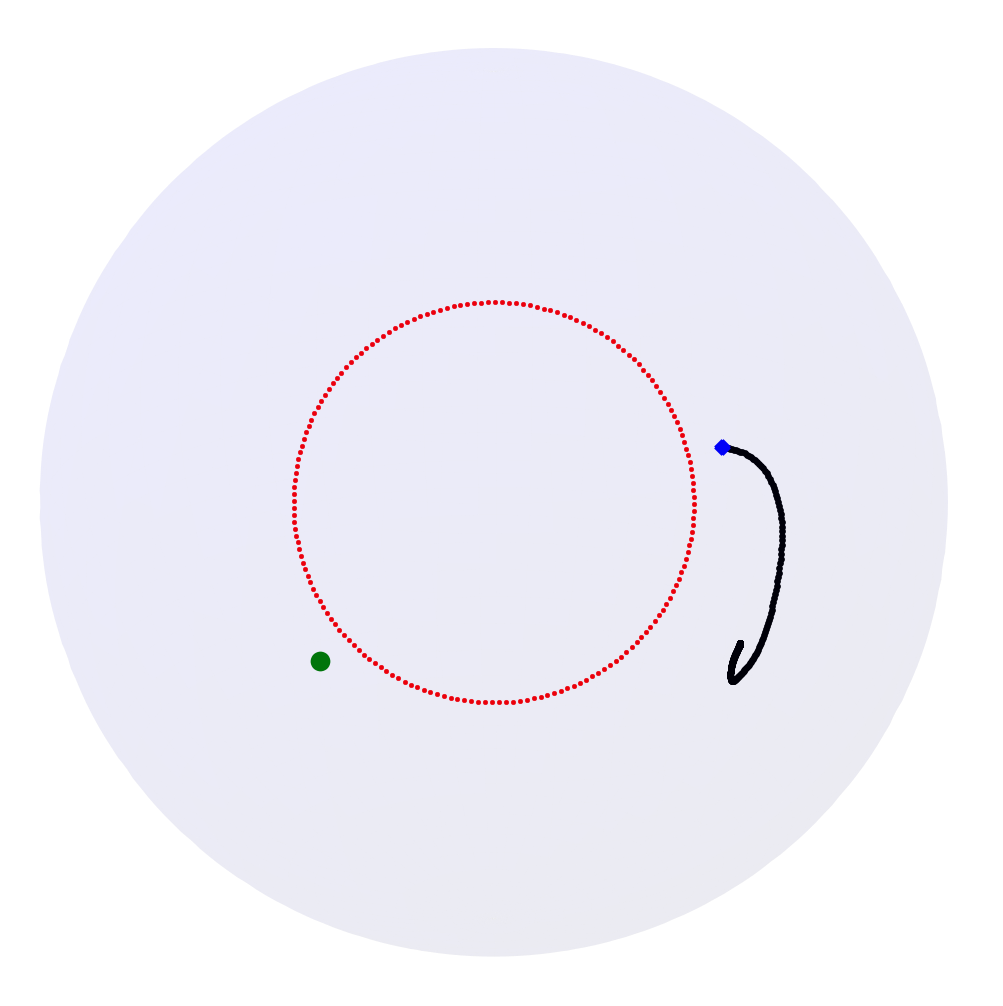}
\caption{Trace of boresight vector on unit sphere}
\end{subfigure}
\begin{subfigure}{0.55\textwidth}
\includegraphics[width=\linewidth]{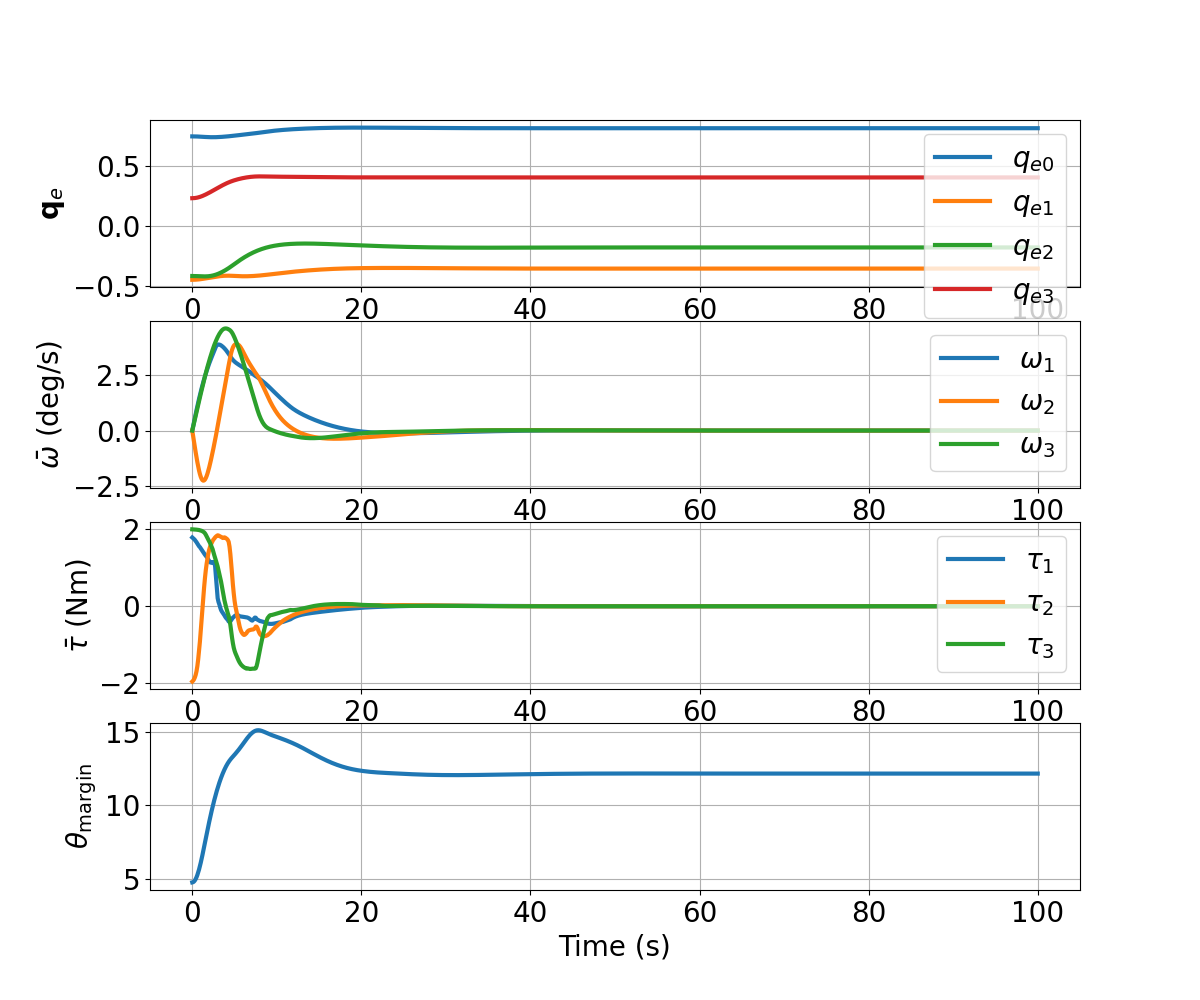}
\caption{Time history of relative attitude, angular velocity, control torque, and $\theta_\text{margin}$}
\end{subfigure}
\caption{Example of non-settled cases (bias)} 
\label{fig:example_non-settled_bias}
\end{figure}

\begin{figure}[htbp]
\begin{subfigure}{0.45\textwidth}
\includegraphics[width=\linewidth]{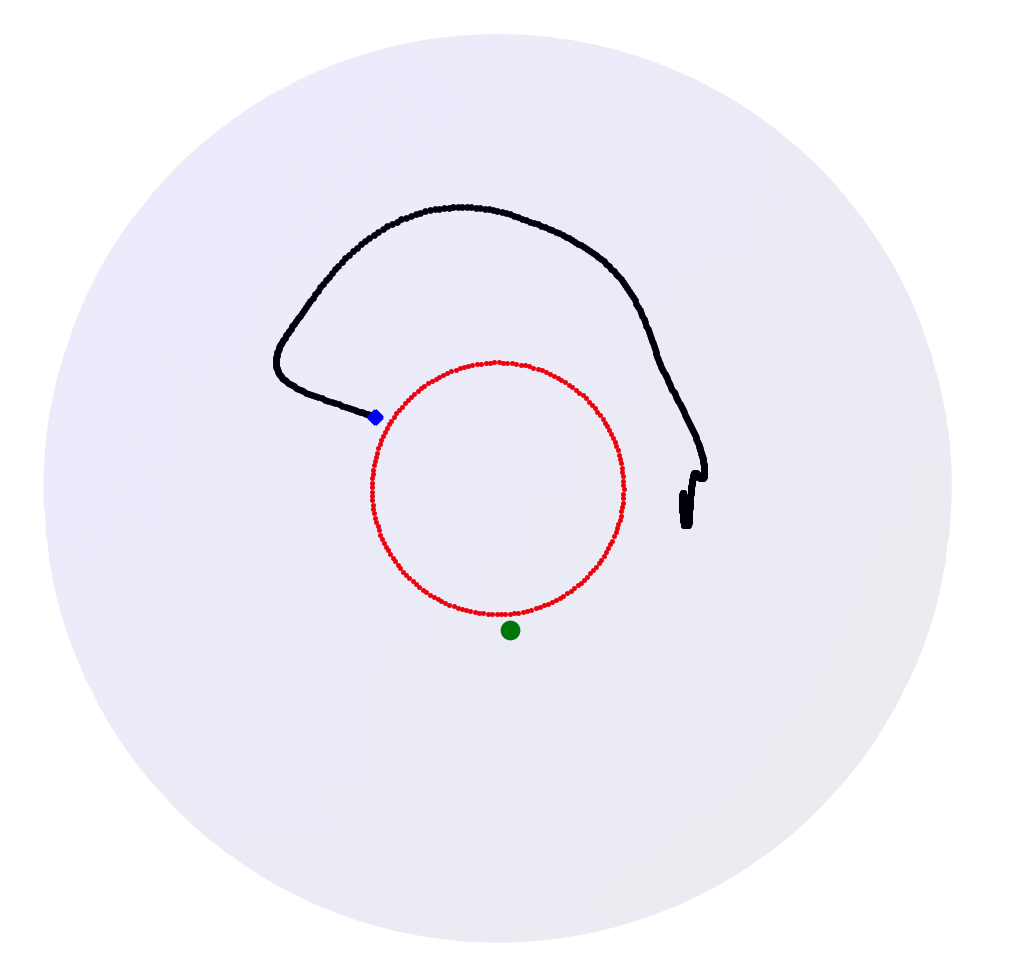}
\caption{Trace of boresight vector on unit sphere}
\end{subfigure}
\begin{subfigure}{0.55\textwidth}
\includegraphics[width=\linewidth]{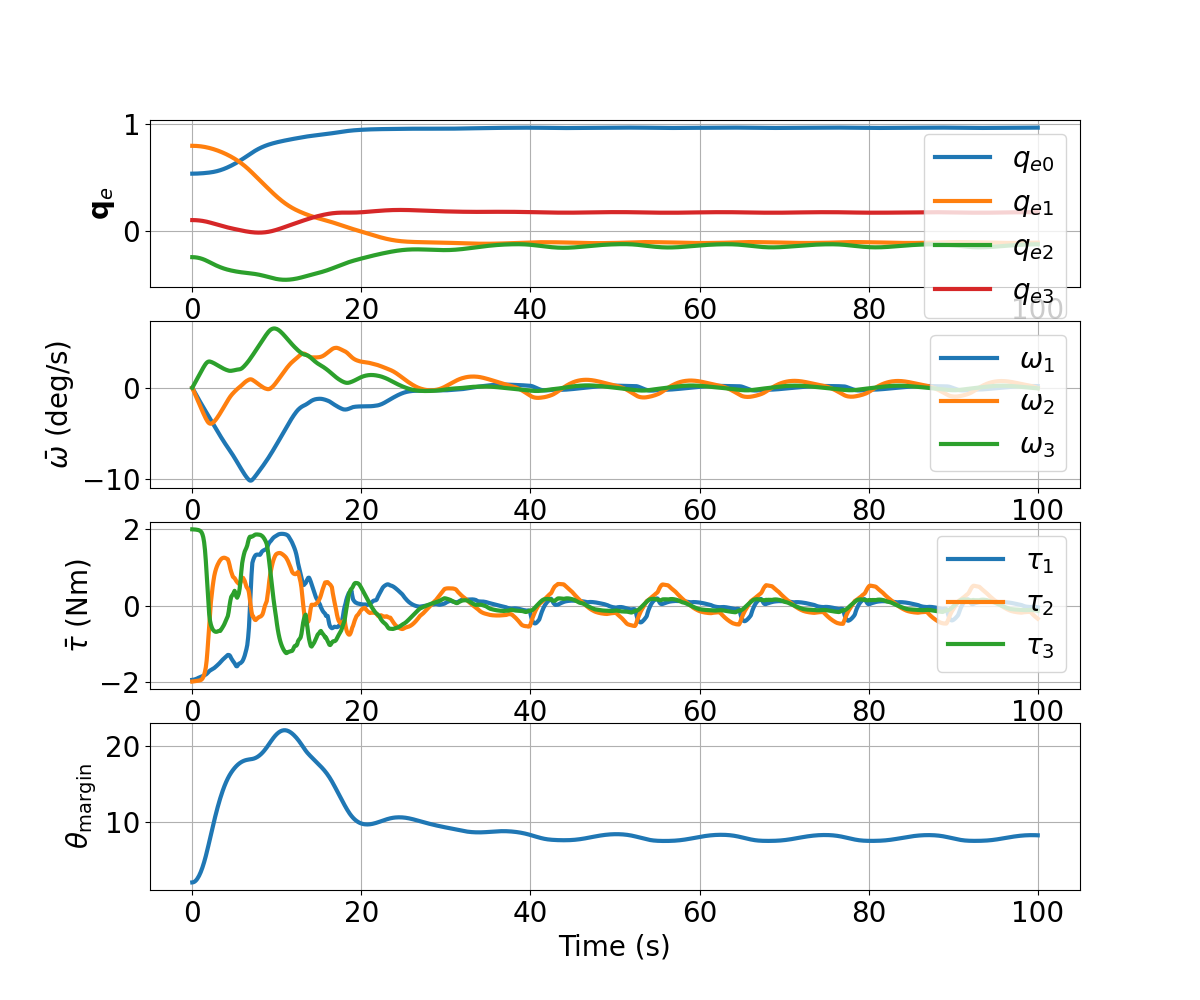}
\caption{Time history of relative attitude, angular velocity, control torque, and $\theta_\text{margin}$}
\end{subfigure}
\caption{Example of non-settled cases (limit cycle)} 
\label{fig:example_non-settled_limit_cycle}
\end{figure}

The Monte Carlo simulation results using the safe-RL framework are summarized in Fig.~\ref{fig:plot_monte_carlo_withSF}. In contrast to the case using the standard RL framework, the violation of the keep-out zone is totally avoided with the safety filter in the loop. However, there are still non-settled cases accounting for $0.22\%$. \textcolor{black}{These non-settled cases fall into two categories: simple biases and limit cycles around certain pointing other than the target pointing. Figures~\ref{fig:example_non-settled_bias} and~\ref{fig:example_non-settled_limit_cycle} show examples for both categories with detailed plots. Individual check of the non-settled cases show that they are caused by the agent itself instead of the safety filter, which implies that the agent from Phase 2 needs further training.}

Table~\ref{tab:monte_carlo_statistics_comparison} (3rd column) shows the detailed results of evaluation metrics. By comparing the 2nd and the 3rd columns in Table~\ref{tab:monte_carlo_statistics_comparison}, it is observed that \textcolor{black}{the mean reward only reduces slightly from $7281.91$ to $7232.96$ and the mean settling time increases slightly from $27.81$ s to $28.47$ s. The mean control effort reduces (from $76.02~\text{N}^2\text{m}^2\text{s}$ to $73.31~\text{N}^2\text{m}^2\text{s}$) and the mean control accuracy remains the same as $0.08$~deg}. 

\begin{figure}[htbp]
\centering
\begin{subfigure}{0.45\textwidth}
\includegraphics[width=\linewidth]{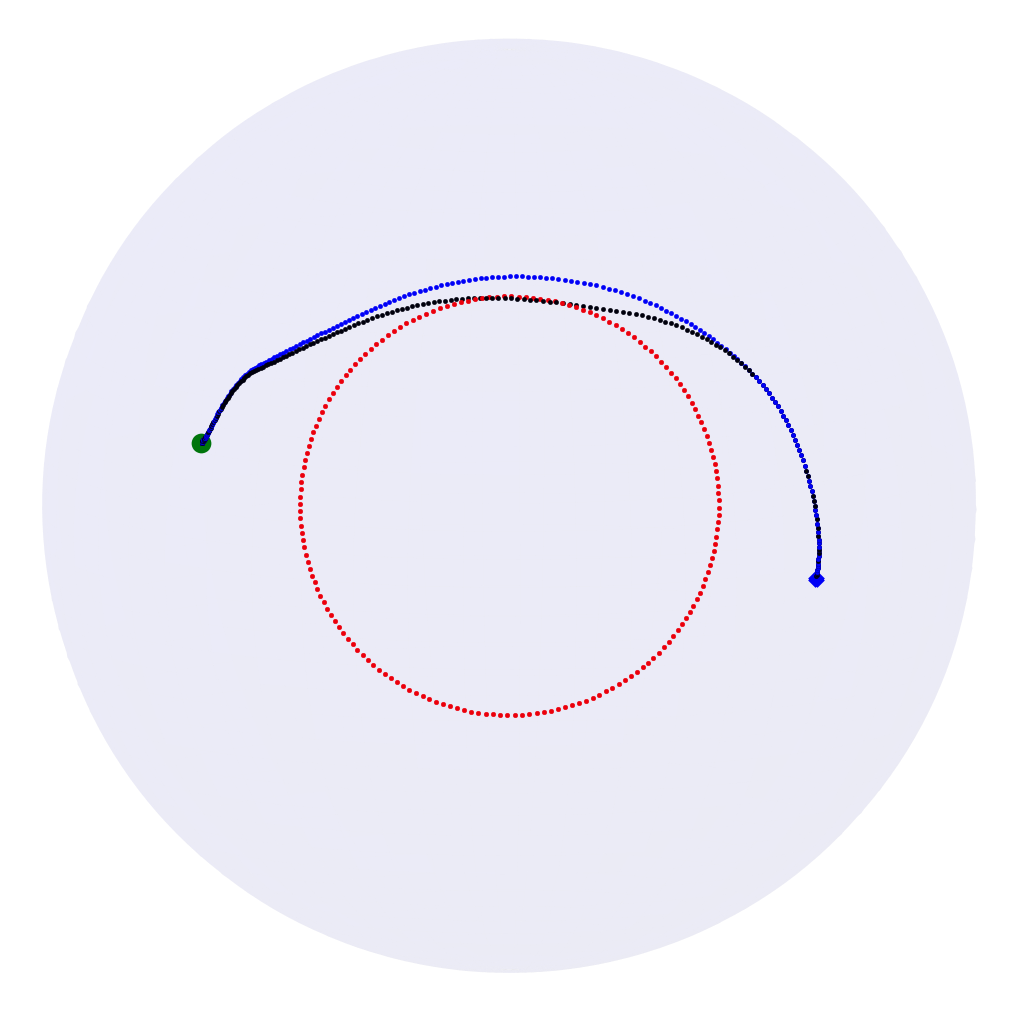}
\caption{Traces of boresight vector on unit sphere (black: without SF; blue: with SF, $\mu=0.0025$)}
\label{fig:plot_3d_w_wo_SFs}
\end{subfigure}
\begin{subfigure}{0.5\textwidth}
\includegraphics[width=\linewidth]{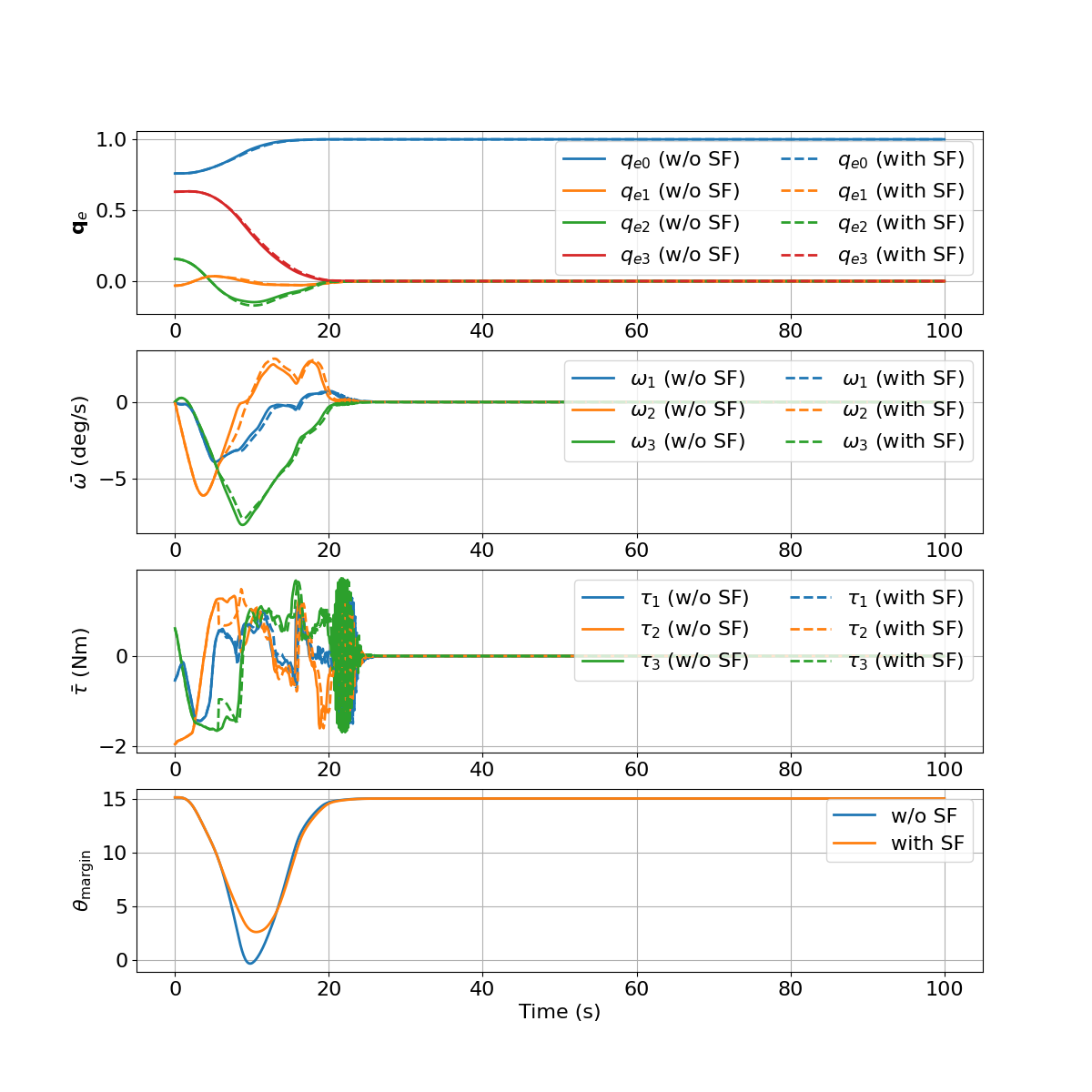}
\caption{Time history of relative attitude, angular velocity, control torque, and $\theta_\text{margin}$}
\label{fig:plot_time_history_w_wo_SFs}
\end{subfigure}
\caption{Example results under agent with and without safety filter in the loop} 
\label{fig:example_result_comparison}
\end{figure}

Figure~\ref{fig:example_result_comparison} presents simulation examples illustrating how the safety filter changes the reorientation maneuver to ensure the keep-out zone avoidance. \textcolor{black}{The black trace in Fig.~\ref{fig:plot_3d_w_wo_SFs} is from a simulation without the safety filter and the blue from a simulation with the safety filter ($\mu = 0.0025$). Figure~\ref{fig:plot_time_history_w_wo_SFs} compares the time history of related data. It is observed that the safety filter with $\mu = 0.0025$ starts to change the maneuver only when the boresight vector approaches the keep-out zone to a certain angular distance.}

\textcolor{black}{Simulations with different parameter settings for the CBF-based safety filter were performed.  It turns out that the parameter $\mu$ has the most significant influence on the performance of the safety filter. Using a small $\mu$ that is not set suitably can lead to worse performance than that in cases without using a safety filter. The 4th column of Table~\ref{tab:monte_carlo_statistics_comparison} presents the result when using a safety filter with $\mu = 0.0001$, which is worse than that without a safety filter in terms of the mean reward, the mean settling time, and the rate of non-settled samples, even though the violation of the keep-out zone is avoided.}  

\section{Conclusions}
This paper proposed a safe DRL solution for spacecraft reorientation subject to a single pointing keep-out zone. The approach is characterized by a novel state space representation, a constraint-aware reward function, and training with the SAC algorithm and curriculum learning. To ensure operational safety, a safety filter was integrated into the action loop, implementing a safe reinforcement learning strategy. The effectiveness of the state space design and the curriculum learning was validated through simulations. Monte Carlo simulations underscore that reward shaping alone cannot reliably prevent constraint violations. In contrast, the framework incorporating a safety filter successfully guarantee the constraint compliance. \textcolor{black}{However, it is also observed that the agent from Phase 2 training may result in non-settled results under certain initial conditions. Further training of the agent is required to address this issue. The current work considers only one keep-out zone for a boresight vector. Future work will extend the current algorithm to account for multiple constraint zones by using long short-term memory (LSTM) networks to determine a single equivalent keep-out zone based on the configuration of the multiple keep-out zones and the current status of the spacecraft. Future work will also involve training and evaluating the agent using a high-fidelity simulation environment (e.g., Basilisk).}

\section*{Appendix}
\label{sec:appendix}
\section*{Acknowledgments}
This work was supported by JMU Seed Grant from University of Würzburg.

\section*{Declaration of Use of Artificial Intelligence}
\label{sec:AI_use_declaration}
Deep reinforcement learning was used for spacecraft attitude control with a single pointing keep-out zone.

\bibliography{CEAS_EuroGNC_references}

\end{document}